\documentclass[conference]{IEEEtran}
\IEEEoverridecommandlockouts
\usepackage{cite}
\usepackage{amsmath,amssymb,amsfonts}
\usepackage{algorithmic}
\usepackage{graphicx}
\usepackage{textcomp}
\usepackage{xcolor}
\usepackage{booktabs} 
\usepackage{algorithm}
\usepackage{algorithmic}
\usepackage{multirow}
\usepackage{amsmath}
\usepackage{amssymb}
\usepackage{subfigure}
\usepackage{epstopdf}
\usepackage{balance}

\newtheorem{definition}{Definition}

\def\BibTeX{{\rm B\kern-.05em{\sc i\kern-.025em b}\kern-.08em
    T\kern-.1667em\lower.7ex\hbox{E}\kern-.125emX}}
\begin{document}

\title{Unsupervised Adversarial Graph Alignment with Graph Embedding}
\author{Chaoqi Chen$^{1}$, Weiping Xie$^{1}$, Tingyang Xu$^{2}$, Yu Rong$^2$, Wenbing Huang$^2$, Xinghao Ding$^1$,\\
 Yue Huang$^{1}$, Junzhou Huang$^2$\\
 {$^1$~Xiamen University, $^2$~Tencent AI Lab}
}

\maketitle

\begin{abstract}
Graph alignment, also known as network alignment, is a fundamental task in social network analysis. Many recent works have relied on partially labeled cross-graph node correspondences, i.e., anchor links. However, due to the privacy and security issue, the manual labeling of anchor links for diverse scenarios may be prohibitive. Aligning two graphs without any anchor links is a crucial and challenging task. In this paper, we propose an Unsupervised Adversarial Graph Alignment (UAGA) framework to learn a cross-graph alignment between two embedding spaces of different graphs in a fully unsupervised fashion (\emph{i.e.,} no existing anchor links and no users' personal profile or attribute information is available). The proposed framework learns the embedding spaces of each graph, and then attempts to align the two spaces via adversarial training, followed by a refinement procedure. We further extend our UAGA method to incremental UAGA (iUAGA) that iteratively reveals the unobserved user links based on the pseudo anchor links. This can be used to further improve both the embedding quality and the alignment accuracy. Moreover, the proposed methods will benefit some real-world applications, \emph{e.g.,} link prediction in social networks. Comprehensive experiments on real-world data demonstrate the effectiveness of our proposed approaches UAGA and iUAGA for unsupervised graph alignment.
\end{abstract}

\begin{IEEEkeywords}
Graph Alignment, Unsupervised, Adversarial, Incremental, Embedding Approach
\end{IEEEkeywords}

\section{Introduction}
\label{sec:Introduction}
With the prosperity of online social networks, people nowadays are usually involved in different social networks simultaneously to enjoy more applications. This provides an opportunity and a challenge to combine heterogeneous social networks and to collectively mine them. More often than not, finding node correspondence across different networks is the very first step for many data mining task and a fundamental building block in numerous applications, such as, the social network link prediction~\cite{dong2012link,zhang2014meta} and the cross-domain recommendation~\cite{hu2013personalized,li2014matching}. The problem formulated thus far is typically referred to as social networks alignment. The social networks can virtually be model as graphs~\cite{yacsar2018iterative,huang2018adaptive,li2019semi}, so the focus of this paper can be stated as aligning two social graphs (also known as social networks). A naive approach for graph alignment is to make use of usernames to identify the same users across different social graphs. However, due to the fact that there exist many users who deliberately use different usernames, such an assumption could be easily violated in most real-world applications.

Most prior efforts, which focus on leveraging the graph structure (\emph{i.e.} topology) for graph alignment, can fall into two main categories. The first category is the \emph{unsupervised} approach, where no information about the node correspondence across graphs is available. These approaches deal with the graph alignment problem by finding the structural similarities between nodes across different graphs. For example, IsoRank~\cite{singh2008global} propagates pairwise topological similarities in the product graph. NetAlign~\cite{bayati2009algorithms} leverages the max-product belief propagation based on the graph topology. However, these approaches are limited to graphs with similar structure and fail to scale up to highly irregular and large graphs. The justification is that such structural similarity is hard to satisfy in some applications and may lead to sub-optimal results. For instance, a user might be very active on one social graph site (\emph{e.g.,} Facebook), but behaves more quietly on another site (\emph{e.g.,} LinkedIn)~\cite{koutra2013big,zhang2016final}.

To alleviate the aforementioned limitations, the second category is the \emph{supervised} approach, where the node correspondences between different graphs are partially available. 
Most of them~\cite{kong2013inferring,zafarani2013connecting,tan2014mapping,zhang2014meta,liu2016aligning} directly utilize the structural features of social graphs (\emph{e.g.,} users' point-to-point constraints, potential manifold alignment, follower-followee relations) to train a classifier under the supervision of known node correspondences. Nevertheless, these approaches may struggle to capture the intrinsic structural information of social graphs and be sensitive to the slight changes or noises of graph structure. Hence, some recent studies~\cite{zhang2016final,yacsar2018iterative} started to consider either leveraging vertex and edge attributes (\emph{e.g.,} demographic information of the users and the communication types between different users) to guide the global structure-based graph alignment or using an embedding approach to capture the structural regularities of social graphs~\cite{man2016predict}, which can further compensate the lack of supervision information and achieve a more accurate alignment.

Despite their efficacy, existing graph alignment methods may be confined by three key bottlenecks. First, the supervised approaches require massive amounts of cross-graph node correspondences and/or attribute information on nodes and edges, however, manual labeling of sufficient training data for diverse social applications on-the-fly is often prohibitive. Second, although the unsupervised approaches sound appealing, their performance is significantly worsen than that of supervised approaches. Third, although there are some methods consider first deriving embeddings for each graph and then learning a transformation between the embeddings, their results are often unsatisfying because their embeddings are predefined. 

In this paper, we introduce a scalable Unsupervised Adversarial Graph Alignment (UAGA) framework that significantly outperforms state-of-the-art unsupervised approaches and reaches or outperforms supervised baselines, without required any cross-graphs annotated data. We only use structure information from two different graphs, which are denoted as the source and target graphs respectively.
The proposed framework consists of three steps, namely, embedding, matching and refining. First, it utilizes unsupervised graph embedding approach to learn the individual feature representations of source and target graphs respectively. Next, it leverages adversarial training to learn a linear mapping from the source embedding space to the target embedding space, 
followed by a refinement procedure which derives the pseudo node correspondences from the resulting shared embedding space and fine-tune the mapping with the closed-form Procrustes solution from~\cite{schonemann1966generalized}. Finally, we extend the UAGA method to incremental UAGA (iUAGA) for improving both the embedding quality and the alignment performance iteratively.

Overall our paper makes the following contributions:
\begin{itemize}
\item{
To the best of our knowledge, we are the first to approach graph alignment in a fully unsupervised manner, where neither anchor links nor node attribute information (\emph{e.g.} user's personal profiles) are demanded. Our UAGA, instead of aligning the nodes between two graphs directly, first learns the node representations of the source and target graphs and then aligns the corresponding embedding spaces using adversarial training in conjunction with an ad-hoc refinement procedure.}

\item{ We highlight the \emph{hubness problem} that a certain number of nodes become hubs and nearest neighbors of many other nodes in a  high-dimensional embedding space, and which, if not be addressed, will make the alignment by nearest searching more difficult and inaccurate. We hence propose a novel cross-graph similarity adaptation method that mitigates such issue and permits a remarkable performance improvement.
}

\item{

To improve the embedding quality, we further propose iUAGA which utilizes the pseudo anchor links obtained in the refinement procedure to extend the training graphs by iteratively revealing the unobserved user links and re-training our UAGA model with the extended training graphs.} 

\item{Comprehensive experiments on real-world data clearly demonstrate the effectiveness of our  proposed approaches for unsupervised graph alignment, compared to the state-of-the-art unsupervised counterparts.}
\end{itemize}

The remainder of the paper is organized as follows. We discuss related work in Section~\ref{sec:Related Work} and introduce preliminary concepts in Section~\ref{sec:preliminary}. Section~\ref{sec:method} introduces the proposed UAGA, including domain-adversarial training, refinement procedure and orthogonality constraint. In Section~\ref{sec:iUAGA} we extend the UAGA to incremental UAGA. Section~\ref{sec:experiments} provides the empirical studies on several standard graph alignment datasets and the analytical experiments will also be implemented.  We conclude the paper in Section~\ref{sec:conclusions}.
\begin{figure*}[!t]
\centering
\includegraphics[width=13cm]{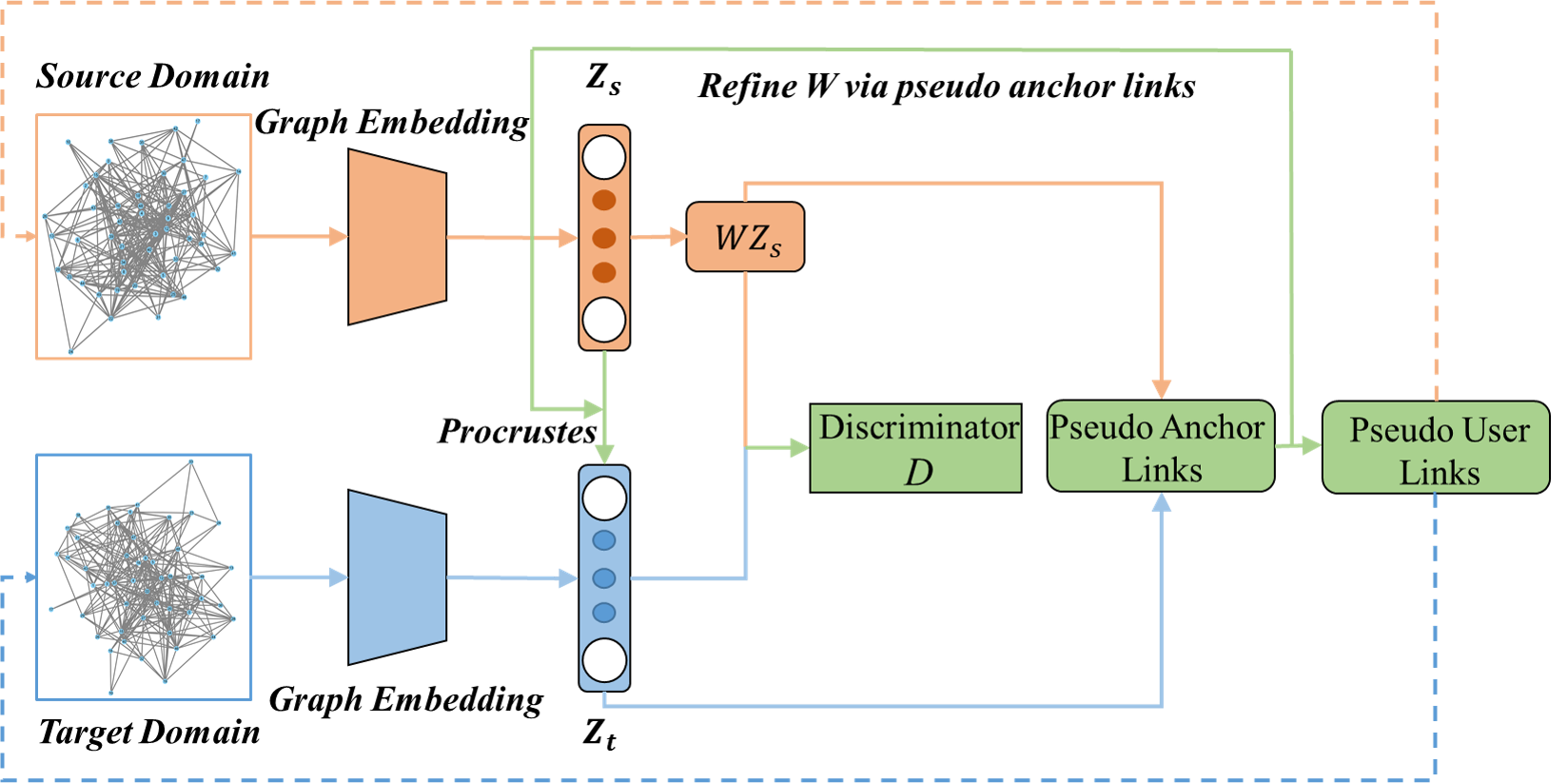}
\caption{The architectures of the proposed UAGA and iUAGA models. $Z_s$ and $Z_t$ denote the source and target embeddings, respectively. In UAGA, the mapping $W$ is refined by Procrustes with pseudo anchor links. In iUAGA, we utilize the derived pseudo anchor links to generate pseudo user links, in order to extend the source and target graphs (the dash lines). And then we iteratively use the extended source and target graphs to improve both the graph embedding and the mapping quality. }\label{fig1}
\end{figure*}

\section{Related Work}
\label{sec:Related Work}
The graph alignment, which aims to find node correspondence across different graphs, has attracted lots of research interests with extensive literatures in the past decade. Graph alignment was first studied by exploring the graph structure to align the topology of the underlying graphs. For example, IsoRank~\cite{singh2008global} aligns different graphs greedily based on the pairwise topology similarities (\emph{i.e.,} node similarity) calculated with spectral graph theory and finds the alignment by solving the maximum bipartite matching. NetAlign~\cite{bayati2009algorithms} proposes a distributed and efficient algorithm for the sparse bipartite graph matching problem by using the max-product belief propagation based on the graph topology. UniAlign~\cite{koutra2013big} proposes to distinguish the nodes of graphs by types (\emph{e.g.} users and groups) and find the correspondences at different granularities at once for the fast bipartite or unipartite graph matching problem. Klau~\cite{klau2009new} formulates the pairwise global graph alignment problem as the graph-based maximum structural matching problem, which is solved by a Lagrangian relaxation approach. These approaches, however, ignore the rich attribute information on nodes and/or edges in many real graphs and are prone to lead to sub-optimal results.

Recently, many attributed-graph-based approaches~\cite{liu2014hydra} have been proposed to tackle the graph alignment problem by leveraging the nodes' attributes (\emph{e.g.} user's social, spatial, temporal and text information). \cite{kong2013inferring} aims to infer the anchor links between two attributed social graphs with a supervised learning model and explicitly ensure the point-to-point constraint by the stable matching. \cite{liu2016aligning} propose to utilize the graph representation learning approach to learn the embeddings of multiple graphs, and the follower-followee relations of each user are represented as the input and output context vectors in the embedded space to enhance the alignment performance. \cite{zhong2012comsoc} models users' interests as latent topics and integrate relationships adaptively by a hierarchical Bayesian model and utilizes transfer learning to transfer both the topological and topical knowledge across subgraphs in a composite graph. \cite{zafarani2013connecting} is an users' behavior-driven approach, which leverages the users' unique behavioral patterns to reduce redundant information so as to construct the users-pecific features for the  classification of corresponding nodes. FINAL~\cite{zhang2016final} holds the view that the topology consistency among different graphs be easily violated in some cases (\emph{e.g.} the active level of the same user in different social graphs may be distinct). Therefore, they leverage the node/edge attribute information to guide the topology-based alignment process based on the alignment consistency principle. gsaNA~\cite{yacsar2018iterative} proposes to align large and highly irregular graphs based on the observation that most prior efforts are limited since they can only handle small graphs (\emph{e.g.} protein protein interaction graphs). Specifically, they reduce the search space of the problem by partitioning the vertices into buckets and present an iterative algorithm to incrementally map vertices of input graphs based on the similarity of the vertices in the same bucket of different graphs. In addition, \cite{zhang2015multiple} proposes to tackle the unsupervised multiple anonymized social graphs alignment problem by penalizing the violation to the the``transitivity law'' and the ``point-to-point property''.

\section{Preliminary}
\label{sec:preliminary}
\subsection{Task Formulation}
We consider the problem of aligning two different social graphs. Denote a social graph as $G = (V,E)$, where $V$ represents the user nodes of the graph, and the edge set $E\subseteq(V\times{V})$ denotes the connections. In this paper, we focus on the following settings: many users are simultaneously involved in two different social graphs, forming \emph{anchor links} across the two graphs. Without loss of generality, we refer to one graph as the source graph, and the other as the target graph, denoted with $G_s$ and $G_t$ respectively. For each node in the source graph, we aim to identify, if any, its counterpart in the target graph. Formally, we can formulate the aforementioned problem as the following graph alignment problem, 

\begin{definition}{(Graph Alignment Problem)}
Given two social graphs $G_s = (V_s,E_s)$ and $G_t = (V_t,E_t)$ and without any observed anchor links (node correspondences), it aims to identify hidden anchor links across $G_s$ and $G_t$ by graph alignment.
\end{definition}\label{definition1}
\subsection{Unsupervised Graph Embedding}
\label{sec:embedding}
Recently, graph embedding models~\cite{wang2016structural,qiu2018network,kazemi2018simple}, for instance, DeepWalk~\cite{perozzi2014deepwalk}, LINE~\cite{tang2015line} and Node2Vec~\cite{grover2016node2vec} were developed to be capable of learning continuous feature representations for nodes in graphs. Specifically, DeepWalk is a unsupervised deep embedding method, which uses local information obtained from truncated random walks to learn features that capture the graph structure independent of the labels' distribution. In this work, we utilizes DeepWalk as an example to learn the source and target graph embeddings, denoted with $Z_s$ and $Z_t$ respectively, where $Z\in{\mathbb{R}}^{|V|\times{d}}$ is a $d$-dimensional feature vector. These embedded feature representations are distributed, which means that each social node $v_i$ is represented by a subset of the latent dimensions. Using these structural feature representations, we are capable of finding the cross-graph node correspondences by learning a mapping function from the source to target embedded feature spaces.

DeepWalk, inspired by Word2Vec~\cite{mikolov2013distributed}, borrows the methodology of Skip-grams~\cite{mikolov2013efficient} for training. The objective of Skip-grams is for each word $w^{(n)}$ in a text corpus, the model is trained to maximize the probability of words $\{w^{(n-k)},...,\\w^{(n-1)},w^{(n+1)},...,w^{(n+k)}\}$ within a window of size $k$ given $w^{(n)}$.
In order to learn the feature representation that captures the shared similarities in local graph structure between nodes (\emph{i.e.,} encoding co-citation similarity), we have the following optimization problem,
\begin{equation}\label{eq1}
\min\limits_{\Phi}\;-\log\Pr(\{v_{i-w},...,v_{i+w}\} \backslash{v_i|\Phi(v_i)}),
\end{equation}
where $w$ is the window size and $\Phi$ is a mapping function: $v\in{V}\mapsto{Z\in{\mathbb{R}}^{|V|\times{d}}}$. Readers can refer to~\cite{perozzi2014deepwalk} for more implementation details.
%
\section{Unsupervised Adversarial Graph Alignment}
\label{sec:method}
The overview of proposed UAGA is shown in Figure~\ref{fig1}. Firstly, we learn the graph feature representations with unsupervised graph embedding approach (\emph{i.e.} DeepWalk). Secondly, we utilize the domain-adversarial training to learn an initial proxy of $W$. Next, we leverage the learned $W$ to select pseudo anchor links for mapping refinement. Finally, we use the refined mapping $W$ to map all source nodes into target latent feature space. We improve the performance over graph alignment by changing the metric of the space, which leads to spread more of those nodes in data-dense areas.

Suppose we have two graph embedding spaces (denoted by the source and the target domains) that trained on two different social graphs. In this paper, we propose to learn a mapping, without requiring any form of cross-domain supervision, between them such that the two different embedding spaces can be well aligned.

Let $Z_s=\{z_s^1,...,z_s^n\}$ and $Z_t=\{z_t^1,...,z_t^m\}$ be two sets of $n$ and $m$ graph embeddings coming from the source and target embedding spaces, respectively. Ideally, if we have known cross-graph node correspondences that specifies which $z_s^i\in{Z_s}$ corresponds to which $z_t^i\in{Z_t}$, we are able to learn a linear mapping $W$ of size $d\times{d}$ between the source and the target space such that
\begin{equation}\label{eq5}
W^* = \mathop{\arg\min}\limits_{W\in{\mathbb{R}^{d^2}}}{\left\Vert{WX-Y}\right\Vert}^2
\end{equation}
where $d$ is the dimension of the embeddings, and $X$ and $Y$ are two aligned matrices of size $d\times{k}$ formed by $k$ node embeddings selected from $Z_s$ and $Z_t$, respectively. During the testing phase, the transformation result of any embeddings $z_s^i$ in the source domain can be defined as $\mathop{\arg\max}_{z_t^j\in{Z_t}}cos(Wz_s^i,z_t^j)$. \emph{Note that the Equation~\eqref{eq5} can directly be used as portion of loss for supervised learning if we have ground-truth node correspondences.}

We show how to learn this mapping $W$ without using any cross-domain supervision. The proposed UAGA framework consists of two steps: domain-adversarial training for learning an initial proxy of $W$, followed by a refinement procedure which utilizes the best matching nodes to create pseudo anchor links for applying Equation~\eqref{eq5}. We will describe the details of each of these steps as follows.
\subsection{Domain-Adversarial Training}
In this step, our ultimate goal is to make the mapped $Z_s$ and $Z_t$ indistinguishable. Firstly, we define a discriminator, which aims to discriminate between elements randomly sampled from $WZ_s = \{Wz_s^1,Wz_s^2,..,Wz_s^n\}$ and $Z_t$. The mapping $W$, which can be seen as the generator, is trained to prevent the discriminator from making accurate predictions by making the embedding features $WZ_s$ and $Z_t$ become indistinguishable. Consequently, the domain-adversarial training procedure is a two-player game, where the first player is the discriminator trained to distinguish the source domain from the target domain, and the second player is the generator trained simultaneously to confuse the discriminator.

Given the mapping $W$, the discriminator, parameterized by $\theta_D$, is optimized by minimizing the following objective function:
\begin{equation}\label{eq6}
\begin{split}
\mathcal{L}_D(\theta_D|W)=&-\frac{1}{n}\sum\limits_{i=1}^n\,\log P_{\theta_D}(source=1|Wz_s^i)\\
                          &-\frac{1}{m}\sum\limits_{i=1}^m\,\log P_{\theta_D}(source=0|z_t^j)
\end{split}
\end{equation}
where $P_{\theta_D}(source=1|Wz_s^i)$ is the probability that embedding $z_s^i$ comes from the source embedding space, vice versa. 
Given the discriminator, the mapping $W$ aims to fool the discriminator's ability to precisely predict the original domain of the embeddings by minimizing the following objective function:
\begin{equation}\label{eq7}
\begin{split}
\mathcal{L}_W(W|\theta_D)=&-\frac{1}{n}\sum\limits_{i=1}^n\,\log P_{\theta_D}(source=0|Wz_s^i)\\
                          &-\frac{1}{m}\sum\limits_{i=1}^m\,\log P_{\theta_D}(source=1|z_t^j)
\end{split}
\end{equation}
To train our model, we follow the standard generative adversarial nets (GANs)~\cite{goodfellow2014generative} training procedure, which alternately and iteratively optimizes the discriminator $\theta_D$ and the mapping $W$ to respectively minimize $\mathcal{L}_D$ and $\mathcal{L}_W$.
\subsection{The Hubness Problem and Refinement Procedure}
\label{sec:Refinement}
The domain adversarial training step learns a mapping function $W$ that matches the global source and target embedding spaces, without considering the complex multimode structures underlying the data distributions. In other words, the fine-grained point-to-point constraint (\emph{i.e.} no two edges share a common endpoint) between two sets of nodes was not explicitly enforced in our unsupervised scenario.

To tackle the aforementioned challenge, we propose a refinement procedure to enable fine-grained point-to-point graph alignment with the generated \emph{pseudo anchor links}. The merits of the refinement procedure can be concluded in two aspects. First, we circumvent the potential mode collapse problem (which is a notorious problem in traditional GANs) in the domain-adversarial training step by introducing additional supervision (pseudo anchor links) to enable point-to-point alignment. Second, when selecting the pseudo anchor links, we mitigate the so-called hubness problem (\emph{i.e.} points tending to be nearest neighbors of many points in high-dimensional spaces)~\cite{dinu2014improving} in deciding mutual nearest neighbors by introducing a Cross-Graph Similarity Scaling (CGSS) scheme.

Specifically, we use the $W$ learned in previous domain-adversarial training step as an initial proxy and then build many pseudo anchor links that specifies which $z_s^i\in{Z_s}$ corresponds to which $z_t^i\in{Z_t}$. To obtain high-quality pseudo anchor links, we 
only retain nodes from $Z_s$ and $Z_t$ that are mutual nearest neighbors.
However, the nearest neighbors are usually asymmetric: $z_t$ being a $K$-NN of $z_s$ does not imply that $z_s$ is a $K$-NN of $z_t$. In the embedding feature spaces, this will lead to a phenomenon that is harmful to aligning node embeddings based on a nearest neighbor rule: some nodes in the embedding space, which we call hubs, are more likely to be the nearest neighbors of many other nodes, but the others (called anti-hubs) are not nearest neighbors of any node.

In our work, we propose the CGSS to mitigate the so-called hubness problem. We consider a bi-partite neighborhood graph, where each node of given anchor links is connected to its $K$ nearest neighbors in the other graph. On this bi-partite graph, the neighborhood associated with a mapped source node embedding $Wz_s$ is denoted by $\mathcal{N}_T(Wz_s)$. All $K$ elements of $\mathcal{N}_T(Wz_s)$ are nodes from the target graph. Likewise the neighborhood associated with an embedding $z_t$ of the target graph is represented by $\mathcal{N}_S(Wz_t)$. The mean similarity of a source embedding $z_s$ to its target neighborhood is denoted by,
\begin{equation}\label{eq8}
\notag
r_T(Wz_s)=\frac{1}{K}\sum\limits_{z_t\in{\mathcal{N}_T(Wz_s)}}\cos(Wz_s,z_t),
\end{equation}
where $\cos(.,.)$ denotes the cosine similarity. Similarly, the mean similarity of a target embedding $z_t$ to its source neighborhood is represented by $r_S(z_t)$. We compute all source and target node embeddings with the efficient nearest neighbors algorithm~\cite{johnson2017billion}. Formally, we utilize these similarities to define a cross-domain similarity measure ${\rm CGSS}(.,.)$ between the mapped source embedding and target embedding as,
\begin{equation}\label{eq9}
{\rm CGSS}(Wz_s,z_t)=2\cos(Wz_s,z_t)-r_T(Wz_s)-r_S(z_t)
\end{equation}

The motivation of the CGSS scheme is intuitive: it increases the similarity associated with isolated node embeddings, while it decreases the ones of embeddings lying in dense areas. In other words, it encourages nodes which lie in the low-density regions to be selected as the pseudo anchor links. Formally, the mapping $W$ is further refined via Procrustes: we apply Equation~\eqref{eq5} on these pseudo anchor links to refine $W$.

\begin{figure}[!t]
\centering
\includegraphics[width=8cm]{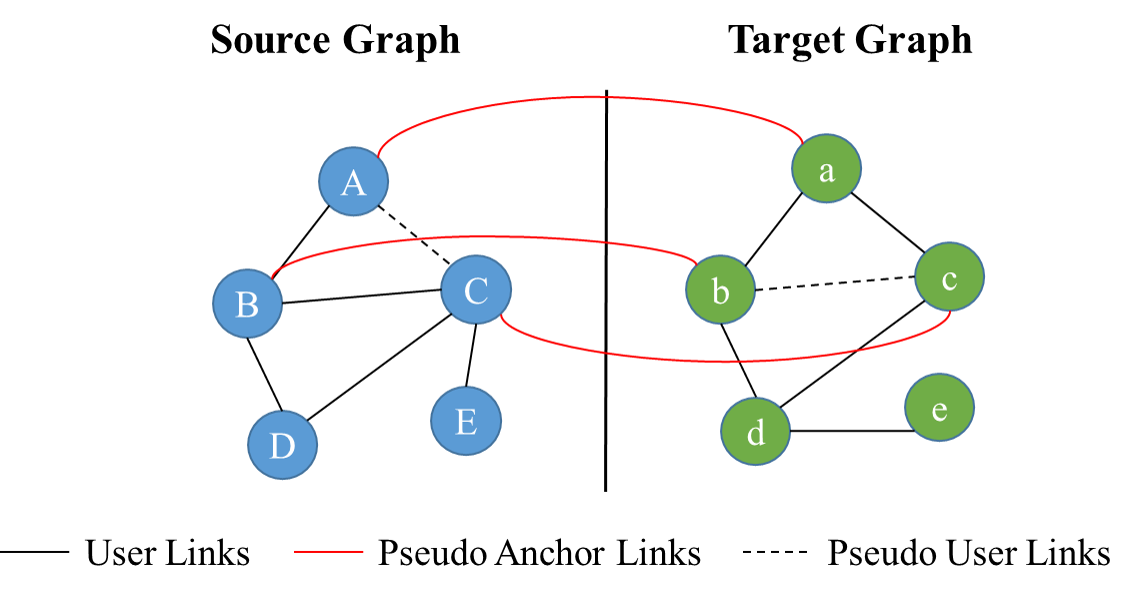}
\caption{The illustration of graph extension by constructing pseudo user links (black dash) with pseudo anchor links (red solid). For simplicity, we utilize the synthetic data for illustration. The user pair (b,c) is not linked in the target graph, but the counterpart user pair (B,C) is linked in the source graph. In addition, we have the pseudo anchor links (B,b) and (C,c). Thus, it is reasonable to add the pseudo user link (b,c) in the target graph. Likewise the user link (A,C) can be added. }\label{fig2}
\end{figure}

\subsection{Orthogonality Constraint}
In our work, we additional propose a simple update step to make the mapping matrix $W$ stays close to an orthogonal matrix as training proceeds\cite{smith2017offline,cisse2017parseval}. The orthogonal constraint has the following advantages: 1) preserves the individual characteristic and quality of source and target feature embeddings respectively; 2) stabilizes the learning process as training proceeds; and 3) preserves the dot product of vectors and their $\ell_2$ distances. Specifically, our model will be updated iteratively and alternatively by using the following update strategy on $W$,
\begin{equation}\label{eq11}
\notag
W\leftarrow{(1+\beta)W-\beta(WW^T)W}
\end{equation}

With the orthogonality constraint, the Equation~\eqref{eq5} can be boiled down to the Procrustes problem, which provides a closed-form solution obtained from the singular value decomposition (SVD) of $YX^T$,
\begin{equation}\label{eq12}
\begin{split}
W^* = &\mathop{\arg\min}\limits_{W\in{\mathbb{R}^{d^2}}}{\left\Vert{WX-Y}\right\Vert}^2=UV^T,\\
&U\sum{V^T}=SVD(YX^T)
\end{split}
\end{equation}

\begin{algorithm}[!t]
\caption{Incremental Unsupervised Adversarial Graph Alignment (iUAGA). $iter$ denotes the iteration of training. The function $DeepWalk$ means the method of DeepWalk. Note that line 2-7 in algorithm is our UAGA method.}\label{algorithm1}
  \begin{algorithmic}[1]
  \REQUIRE $G_s = (V_s,E_s)$, $G_t = (V_t,E_t)$, $\hat{T}=\varnothing$
  \WHILE {not converge}
  \STATE $\{z_s\}_{i=1}^n = DeepWalk(G_s)$ \\
         $\{z_t\}_{j=1}^m = DeepWalk(G_t)$
  \FOR {$j=1$ \textbf{to} $iter$}
  \STATE Run the domain-adversarial training step by alternatively using Equation~\eqref{eq6} and Equation~\eqref{eq7}
  \ENDFOR
  \STATE Acquire a set of pseudo anchor links $\hat{T}$ 
  \STATE Refine the mapping $W$ via Procrustes (\emph{i.e.} Equation~\eqref{eq12}) by using the pseudo anchor links $\hat{T}$
  \STATE Extend both the graphs $G_s$ and $G_t$ into $\hat{G}_s$ and $\hat{G}_t$ by using Equation~\eqref{eq10}
  \STATE $G_s=\hat{G}_s$, $G_t=\hat{G}_t$ and $\hat{T}=\varnothing$
  \ENDWHILE
  \STATE Align $G_s$ and $G_t$ by using the learned mapping $W$ and the distance metric (\emph{i.e.} Equation~\eqref{eq9})
  \end{algorithmic}
\end{algorithm}

\section{Incremental UAGA}
\label{sec:iUAGA}
To further effectively use the pseudo anchor links for learning more accurate mapping, we further propose an incremental UAGA (iUAGA) procedure, which progressively selects a set of pseudo anchor links and then leverage them to refine the mapping $W$ by using Equation~\eqref{eq5}. We select a pseudo anchor link if the following two conditions are satisfied. First, we require the cross-domain similarity (\emph{i.e.} Equation~\eqref{eq9}) should exceed the threshold parameter, which we set as 0.7 or 0.75 in the experiment. The second requirement is that the pairs of embedding features are mutually nearest neighbors of each other according to CGSS. We suppose that unless they are mutual nearest neighbors, the pairwise relationship is not reliable. These two requirements decrease the number of the pseudo anchor links, but improve its accuracy and the final alignment performance. The set of pseudo anchor links is denoted by $\{\hat{T}={(v_s,v_t)|v_s\in{V_s},v_t\in{V_t}}\}$.

Another challenging problem in graph alignment is that some existing edges are still unobserved, as they have neither been explicitly built nor been crawled~\cite{liu2016aligning}. To solve this problem, previous supervised approaches~\cite{man2016predict,liu2016aligning} extend both the source and target graphs with the help of the observed anchor links and the structure of the other graph. In our unsupervised scenario, however, it is very difficult or even impossible to reveal these unobserved edges, which may easily result in unreliable representations when embedding graphs into the latent feature spaces. Motivated by this, our approach proposes to progressively compensate the lack of cross-graph supervision with the refinement procedure by pseudo-labeling many reliable anchor links, which can help to reveal the unobserved edges (\emph{i.e.} intra-graph pseudo user links). The justification is that if two nodes are unconnected in one graph, but their counterparts (according to the pseudo anchor links) are linked in the other graph, it is feasible to add an edge between them in the present graph, as shown in Figure~\ref{fig2}. By leveraging the pseudo-labeled user links, we can further improve both the graph embedding quality and the final alignment performance. The reason behind it is that the extended graphs can provide richer structural information for graph embedding and mapping.

Formally, given two graphs $G_s$ and $G_t$ with pseudo anchor links $\hat{T}$, the extended graph $\hat{G}_s$ of the source graph $G_s$ can be denoted as,
\begin{equation}\label{eq10}
\begin{split}
\hat{V}_s =& V_s\\
\hat{E}_s =& E_s\cup\{(v_s^i,v_s^j):(v_s^i,v_t^k)\in{\hat{T}},\\
&(v_s^j,v_t^l)\in{\hat{T}},(v_t^k,v_t^l)\in{E_t}\}
\end{split}
\end{equation}
Likewise we can extend the target graph $G_t$ into $\hat{G}_t$.

\textbf{Time Complexity Analysis}
The entire procedure of training the iUAGA is shown in Algorithm~\ref{algorithm1}.
The graph embedding by leveraging the DeepWalk can be computed in $O(|V|\log|V|)$~\cite{perozzi2014deepwalk}. Algorithm~\ref{algorithm1} iteratively and progressively generates pseudo anchor links and pseudo user links and linearly maps the source embedding features to the target space, and it can be computed in $O(|V|)$. Therefore, the overall time complexity of iUAGA is $O(|V|^2\log|V|)$.
\section{Experiments}
\label{sec:experiments}
In this section, we extensively evaluate our approach and compare it with both supervised and unsupervised graph alignment baselines on real-world data. We also provide a detailed analysis of the proposed framework, empirically demonstrating the effect of our contributions. All the results in the following are reported as averages over five runs.

\subsection{Experimental Configuration}
\subsubsection{Datasets}
To evaluate the performance thoroughly, we consider different types of real-world graphs with different scales to generate datasets. Table~\ref{table3} shows the statistics of these real-world social graphs~\footnote{http://aminer.org/cosnet}.

\subsubsection{Aligned Graph Generation}
\label{sec:data}
Given a graph $G$, we attempt to generate two subgraphs, which denoted by $G_s$ and $G_t$, for the following evaluations. For example, we are first given a graph $G$~(Last.fm) with 9997 nodes together with their 511490 following edges. Then, we randomly discard 5\% edges of $G$ and treat it as the source subgraph $G_s$, where nodes are all inherited from $G$. Finally, we further discard another (different from $G_s$) 5\% edges of $G$ and treat it as the target subgraph $G_t$. With such a strategy, both $G_s$ and $G_t$ share the common edge ratio $\lambda_e=0.9$ from $G$, which reflects the \emph{overlap} level. For the rest 10\%, randomly discarded 5\% belongs to $G_t$ and the other 5\% belongs to $G_s$, which reflects the \emph{difference} level. Note that only edges can be different between $G_s$ and $G_t$, every pair of node correspondence between $G_s$ and $G_t$ are linked as the ground-truth anchor links. In training phase, we have no access to any anchor links. Specifically, we select the most informative nodes, whose degree is higher than 3, as our $G$ from the original graph.
\begin{center}
\begin{table}[!t]
\caption{The overview of different social graphs.}\label{table3}
\centering
\begin{tabular*}{\hsize}{@{}@{\extracolsep{\fill}}crr@{}}
\toprule
\textbf{Graphs} & \textbf{Nodes} & \textbf{Edges} \\
\hline
  Last.fm & 9997 & 511490 \\
  Flickr & 10000 & 1684207 \\
  MySpace & 10000 & 104056 \\
\bottomrule
\end{tabular*}
\end{table}
\end{center}
\subsubsection{Comparison Methods}
We compare the performance of the proposed UAGA and iUAGA algorithms against four unsupervised \emph{centrality-based} (centrality: computes node centrality measures for a graph) graph alignment algorithms: Degree, Closeness, Betweenness and Eigenvector~\footnote{http://snap.stanford.edu/snap/doc.html}. For the proposed UAGA model, we consider several variants of it in the following experiments:
\begin{itemize}
    \item \textbf{Adv-NN} uses domain-adversarial training to learn the mapping and then leverages the roughly learned mapping to find anchor links according to Nearest Neighbors similarity measure (NN);
    \item \textbf{Adv-CGSS} replaces NN by the proposed CGSS scheme;
    \item \textbf{Adv-Refine-NN} adds the refinement procedure to Adv-NN;
    \item \textbf{Adv-Refine-CGSS} further replaces the NN of Adv-Refine-NN by the CGSS scheme. By adding the incremental training procedure to Adv-Refine-CGSS, we can obtain the iUAGA model.
\end{itemize}

\begin{center}
\begin{table*}[thb]
\vspace{1ex}
\caption{Experimental results on real-world data given different $Precision@N$ settings and $\lambda_e=0.9$.}\label{table1}
\centering
\begin{tabular*}{\hsize}{@{}@{\extracolsep{\fill}}c|cccccccccccc@{}}
\toprule
\multirow{2}{*}{Method} & \multicolumn{3}{c}{Last.fm $\leftrightarrow$ Last.fm} & \multicolumn{3}{c}{Flickr $\leftrightarrow$ Flickr} & \multicolumn{3}{c}{MySpace $\leftrightarrow$ MySpace} & \multicolumn{3}{c}{Avg} \\
\cline{2-13}
& P@1 & P@5 & P@10 & P@1 & P@5 & P@10 & P@1 & P@5 & P@10 & P@1 & P@5 & P@10 \\
\midrule
\midrule
\multicolumn{13}{c}{\emph{Supervised methods with cross-graph supervision}}\\
\midrule
Procrustes-NN & 0.4479 & 0.6467 & 0.8534 & 0.4801 & 0.6554 & 0.8887 & 0.2144 & 0.4289 & 0.5479 & 0.3809 & 0.5770 & 0.7633 \\
Procrustes-CGSS & 0.4836 & 0.6754 & \textbf{0.9211} & 0.5346 & 0.7056 & \textbf{0.9679} & \textbf{0.2880} & \textbf{0.5136} & \textbf{0.6067} & 0.4354 & 0.6316 & \textbf{0.8319} \\
\midrule
\midrule
\multicolumn{13}{c}{\emph{Unsupervised methods without cross-graph supervision}}\\
\midrule
Degree & 0.0167 & 0.0938 & 0.2043 & 0.0124 & 0.1603 & 0.2237 & 0.0266 & 0.1286 & 0.1471 & 0.0557 & 0.1276 & 0.1917 \\
Closeness & 0.0094 & 0.0352 & 0.1178 & 0.0096 & 0.0844 & 0.1657 & 0.0137 & 0.1035 & 0.2035 & 0.0109 & 0.0744 & 0.1623 \\
Betweenness & 0.0076 & 0.0245 & 0.1056 & 0.0084 & 0.0291 & 0.1045 & 0.0081 & 0.0410 & 0.1106 & 0.0080 & 0.0317 & 0.1069 \\
Eigenvector & 0.0105 & 0.1291 & 0.1943 & 0.0101 & 0.0945& 0.1847 & 0.0181 & 0.1125 & 0.2045 & 0.0129 & 0.1120 & 0.1945 \\
\midrule
Adv-NN & 0.3368 & 0.5372 & 0.6234 & 0.3750 & 0.6092 & 0.6970 & 0.1414 & 0.2744 & 0.3466 & 0.2844 & 0.4736 & 0.5557 \\
Adv-CGSS & 0.4361 & 0.6429 & 0.7171 & 0.4625 & 0.6923 & 0.7773 & 0.1858 & 0.3301 & 0.3956 & 0.3615 & 0.5551 & 0.6300 \\
Adv-Refine-NN & 0.4404 & 0.6400 & 0.7128 & 0.5205 & 0.7240 & 0.7909 & 0.1642 & 0.2906 & 0.3984 & 0.3750 & 0.5515 & 0.6340  \\
Adv-Refine-CGSS & 0.5445 & 0.7286 & 0.7802 & 0.5990 & 0.7835 & 0.8448 & 0.2062 & 0.3407 & 0.4389 & 0.4499 & 0.6176 & 0.6880  \\
iUAGA & \textbf{0.6473} & \textbf{0.8179} & 0.8712 & \textbf{0.6343} & \textbf{0.8106} & 0.8743 & 0.2073 & 0.3769 & 0.4815 & \textbf{0.4963} & \textbf{0.6685} & 0.7423 \\
\bottomrule
\end{tabular*}
\end{table*}
\end{center}

We also compare the performance against two \emph{supervised} variants of the proposed UAGA, including \textbf{Procrustes-NN} and \textbf{Procrustes-CGSS}, which utilize 30\% ground-truth anchors links for Procrustes and the rest of anchor links are used for testing. For the sake of fair comparison, we will not utilize the ground-truth anchors links to extend the source and the target user links, which will introduce additional ground-truth user links to the source and target graphs. Note that the focus of this paper is unsupervised setting, so we do not compare with the state-of-the-art supervised methods.
\subsubsection{Evaluation Metric and Scenarios}
To quantitatively evaluate the proposed model, we consider $Precision@N$ as the performance metric. For $Precision@N$, we use the learned mapping $W$ and the similarity measure (NN or CGSS) to output a list of candidate nodes (\emph{i.e.} the top $N$ nodes in descending order according to the similarity values) for each node in the source graph, and then compute the alignment accuracy with respect to the ground-truth.

Based on the aforementioned real-world graphs, we follow the predefined graph generation approach (Section~\ref{sec:data}) to construct three alignment scenarios for evaluations.

\paragraph{S1. Last.fm $\leftrightarrow$ Last.fm} Given a graph with 9997 nodes together with 511490 edges, we generate $G_s$ and $G_t$ respectively.

\paragraph{S2. Flickr $\leftrightarrow$ Flickr} Given a graph with 10000 nodes, together with 1684207 edges, we generate $G_s$ and $G_t$ respectively.

\paragraph{S3. MySpace $\leftrightarrow$ MySpace} Given a graph with 10000 nodes together with 511490 edges, we generate $G_s$ and $G_t$ respectively.

\paragraph{S4. Flickr $\leftrightarrow$ Last.fm} We have a source graph (Last.fm) with 996 nodes, together with 1544 edges. We have a target graph (Flickr) with 1001 nodes, together with 1513 edges. The source and target graphs own 510 ground-truth anchor links (provided by the original datasets).

\paragraph{S5. Flickr $\leftrightarrow$ Myspace} We have a source graph (Flickr) with 1000 nodes, together with 1623 edges. We have a target graph (MySpace) with 997 nodes, together with 1402 edges. The source and target graphs own 509 ground-truth anchor links (provided by the original datasets).

\subsubsection{Model Architectures and Training Details}
The graph embeddings were trained using DeepWalk with Skip-grams by setting the window size to 5. The embedding dimension is set to 32, thus the mapping $W$ has size of $32\times{32}$. For the adversarial training, the discriminator was a two-layer neural network of size $2048$ with Leaky-ReLU as the activation function. The input to the discriminator is corrupted with dropout noise with a rate of 0.1. We include a smoothing coefficient $s = 0.2$ in the discriminator predictions. Both the discriminator and $W$ were trained by stochastic gradient descent (SGD) with a fixed learning rate of $10^{-3}$ and a decay of $0.95$. The batch size was set to 1000 in each experiment.

For the CGSS, we have a hyper-parameter $K$ which defines the size of the neighborhood. We empirically found that the performance is insensitive to $K$, and thus $K$ does not need cross-validation. The results are barely change when $K$= 5, 10 and 50. Consequently, $K$ was set to 10 in all experiments.

In our experiments, we used C++ implementations of Degree, Closeness, Betweenness, Eigenvector algorithms from their source codes~\footnote{https://github.com/snap-stanford/snap}.
Experiments were carried out on machine that has Tesla P40 GPU, 24GB of memory, Intel(R) Xeon(R) CPU E5-2680 v4 $@$2.40GHz 14-core, 2TB disk space, running Ubuntu GNU/Linux. UAGA and iUAGA are implemented by PyTorch.

\subsection{Results and Discussion}
\subsubsection{Graph Alignment Performance}
We first compare the proposed model with \emph{topology-based} baselines, without leveraging the node/edge attribute information to guide the alignment process. The results are presented in Table~\ref{table1}, and we report $Precision@N$ for $N=1,5,10$. The UAGA and iUAGA models are found to significantly and consistently outperform unsupervised comparison methods on all graph alignment scenarios given different experimental settings ($Precision@N$). Moreover, iUAGA is the top-performing unsupervised model and significantly outperforms the unsupervised baselines. Specifically, iUAGA achieves 30\% to 70\% improvements in terms of the alignment accuracy over the existing unsupervised methods.

Meanwhile, the experimental results reveal several insightful observations.
\begin{itemize}
\item Both the UAGA and iUAGA models substantially outperform previous unsupervised alignment approaches (\emph{e.g.} Degree and Eigenvector), which clearly demonstrates the contributions of the proposed method in learning fine-grained one-to-one cross-graph alignment.
\item Both Adv-CGSS and Adv-Refine-CGSS outperform Adv-NN and Adv-Refine-NN. This validates that CGSS provides a strong and robust similarity measure for finding anchor links when generating the pseudo anchor links or executing the testing procedure. Moreover, it also empirically verifies that the CGSS is capable of alleviating the so-called hubness problem and provides reliable pseudo anchor links.
\item Looking at the differences between Adv-Refine-CGSS and Adv-CGSS or Adv-Refine-NN and Adv-NN, we can observe that the proposed refinement procedure provides a significant gain in performance over different scenarios. This indicates that it is beneficial to select pseudo anchor links with the guidance of cross-graph similarity measure.
\item iUAGA significantly outperforms Adv-Refine-CGSS ($Precision@N$: 0.4499 $\rightarrow$ 0.4963, 0.6176 $\rightarrow$ 0.6685, 0.6880 $\rightarrow$ 0.7423), which demonstrates that it is beneficial to iteratively reveal the unobserved user links by the pseudo anchor links, which contributes to refine the roughly learned mapping by using the enlarged training dataset.
\item The proposed iUAGA either is on par, or outperforms supervised baselines, \emph{e.g.,} Avg: \textbf{0.4963}~$\leftrightarrow$~0.4354, \textbf{0.6685}~$\leftrightarrow$~0.6316 and 0.7423~$\leftrightarrow$~\textbf{0.8319}, which further demonstrates that the proposed unsupervised method can even exceed supervised methods without requiring any cross-domain correspondence.
\end{itemize}

\begin{figure}[!t]
\centering
\subfigure[S1: Last.fm-Last.fm]{
\label{fig31}
\includegraphics[width=1.67in]{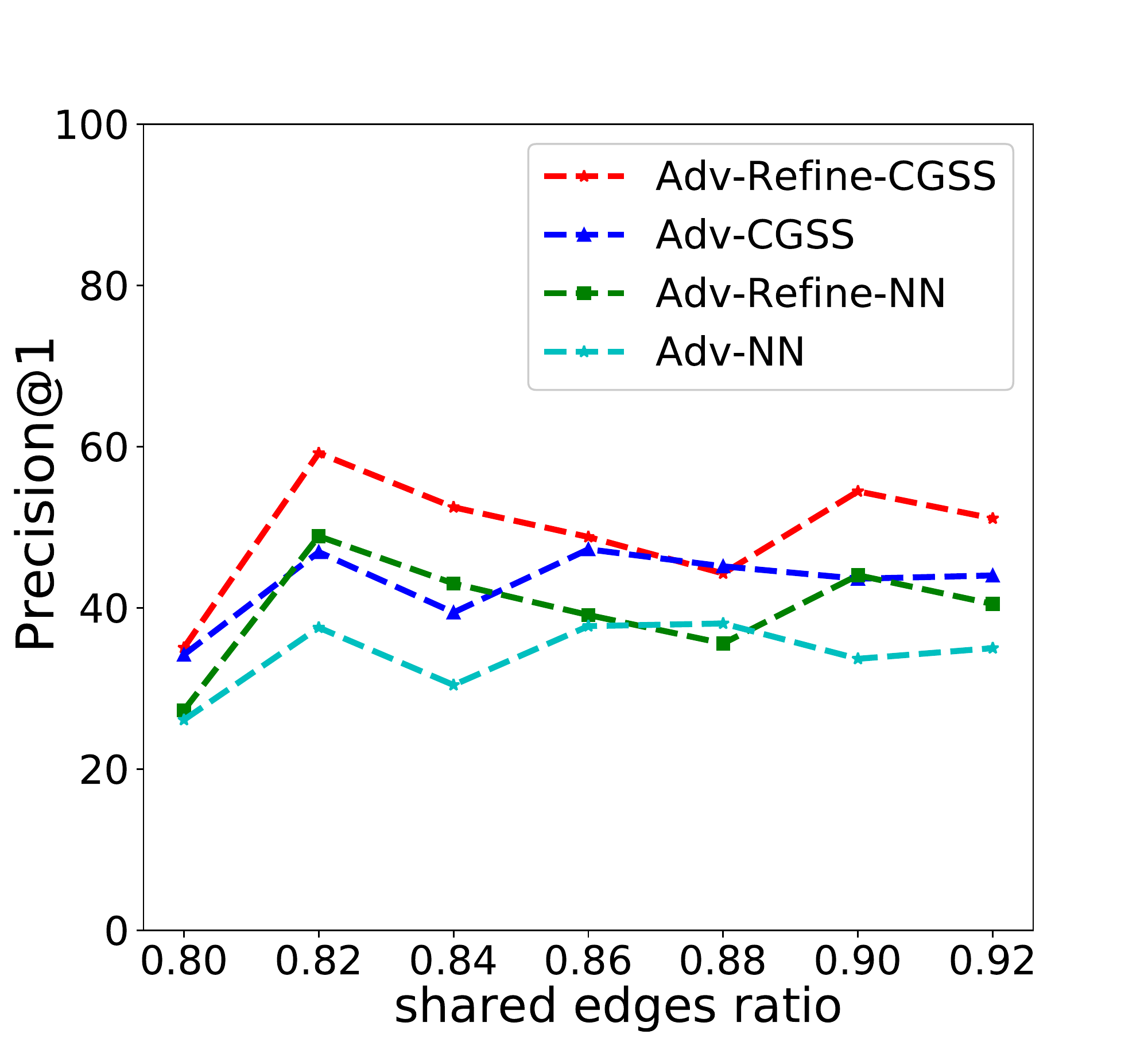}}
\subfigure[S2: Flickr-Flickr]{
\label{fig32}
\includegraphics[width=1.67in]{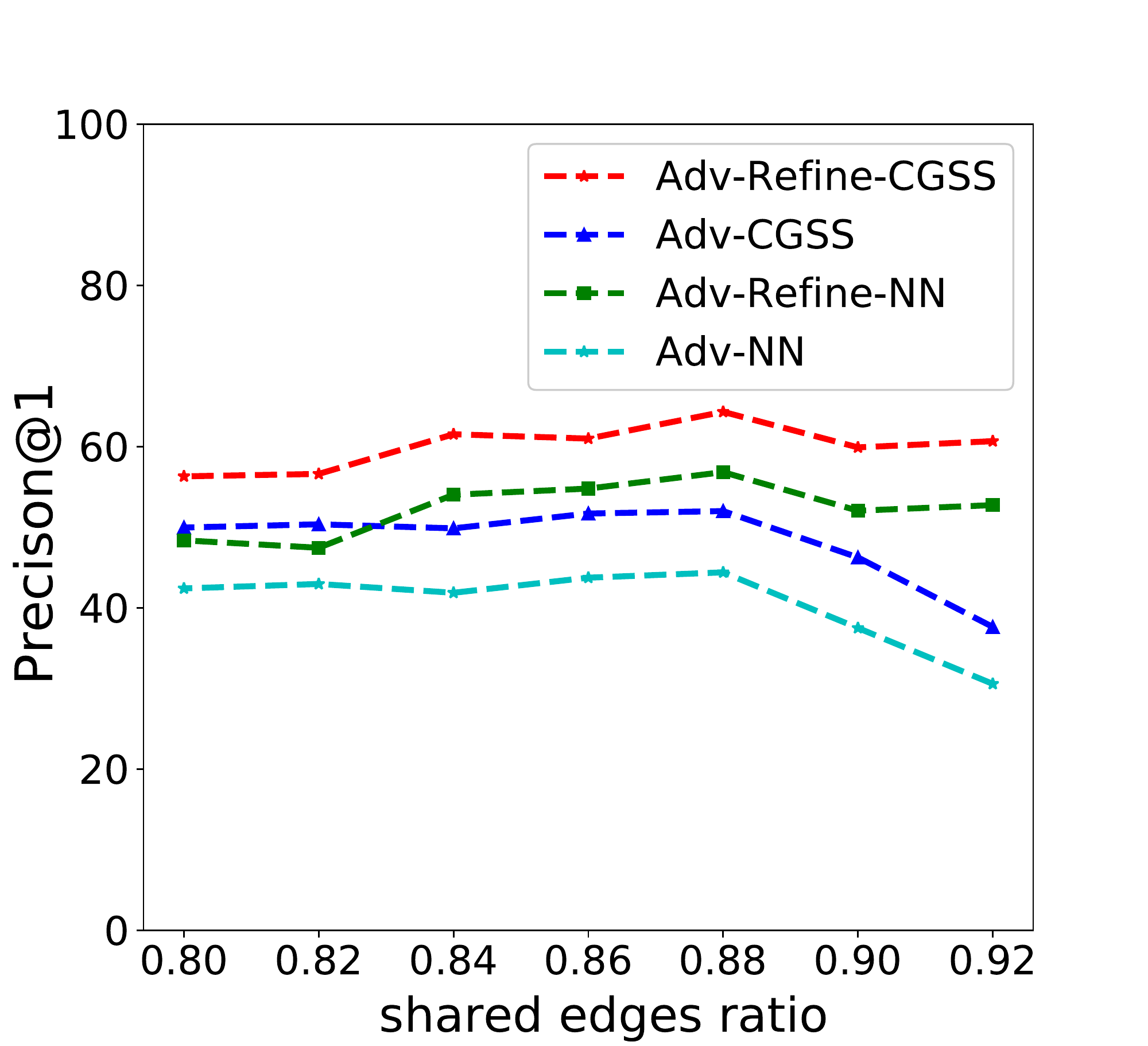}}
\caption{The Precision@1 result for (a) Last.fm-Last.fm graphs, (b) Flickr-Flickr graphs, and $\lambda_e$ ranges from 0.8 to 0.92.}
\label{fig.lambdae}
\end{figure}

\begin{center}
\begin{table}[thbp]
\caption{Experimental results of anchor link inference between different social graphs given different $Precision@N$ settings.}\label{table2}
\centering
\setlength{\tabcolsep}{0mm}{
\begin{tabular*}{\hsize}{@{}@{\extracolsep{\fill}}c|cccccc@{}}
\toprule
\multirow{2}{*}{Method} & \multicolumn{3}{c}{Flickr $\leftrightarrow$ Last.fm} & \multicolumn{3}{c}{Flickr $\leftrightarrow$ MySpace} \\ 
\cline{2-7}
& P@10 & P@20 & P@30 & P@10 & P@20 & P@30 \\
\midrule
\midrule
\multicolumn{7}{c}{\emph{Supervised methods with cross-graph supervision}}\\
\midrule
Procrustes-NN & 0.2764 & 0.3809 & 0.4930 & 0.2905 & 0.3671 & 0.5125 \\
Procrustes-CGSS & 0.3227 & 0.4246 & \textbf{0.5581} & \textbf{0.3505} & \textbf{0.4271} & \textbf{0.5870} \\
\midrule
\midrule
\multicolumn{7}{c}{\emph{Unsupervised methods without cross-graph supervision}}\\
\midrule
Degree & 0.0364 & 0.0821 & 0.1149 & 0.0524 & 0.1073 & 0.1426 \\
Closeness & 0.0265 & 0.0539 & 0.0944 & 0.0291 & 0.0753 & 0.1008 \\
Betweenness & 0.0130 & 0.0452 & 0.0871 & 0.0159 & 0.0565 & 0.0863 \\
Eigenvector & 0.0402 & 0.0977 & 0.1235 & 0.0642 & 0.1137 & 0.1472 \\
\midrule
Adv-NN & 0.1673 & 0.2490 & 0.3222 & 0.1794 & 0.2602 & 0.3587  \\
Adv-CGSS & 0.2483 & 0.3508 & 0.4192 & 0.2595 & 0.3243 & 0.4109 \\
Adv-Refine-NN & 0.2515 & 0.3673 & 0.4310 & 0.2746 & 0.3502 & 0.4371 \\
Adv-Refine-CGSS & \textbf{0.3344} & \textbf{0.4684} & 0.5127 & 0.3209 & 0.4010 & 0.5059 \\
\bottomrule
\end{tabular*}}
\end{table}
\end{center}

\vspace{-4ex}
\subsubsection{Influence of $\lambda_e$}
we evaluate the impact of the number of the cross-graph shared edges (the overlapped edge ratio $\lambda_e$ is ranging from 0.8 to 0.92 and the number of vertices remains unchanged) on the alignment accuracy. The experimental results on $S1$ and $S2$ are demonstrated in Figure~\ref{fig31}-\ref{fig32}. It is noteworthy that the $Precision@1$ is not monotonically increased with the increasing of $\lambda_e$. On the one hand, the justification is that the source and target graphs are vulnerable to contain isomorphic nodes (\emph{i.e.,} nodes in different graphs share the same/analogous topology structure) when the overlapped edge ratio $\lambda_e$ is large. On the other hand, the $Precision@1$ will decease when $\lambda_e$ decreases to a certain small value since the source and target graphs would be much dissimilar in that case.

\begin{center}
\begin{table}[!t]
\caption{precision$@$k on $S1$ for link prediction.}\label{table4}
\centering
\begin{tabular*}{\hsize}{@{}@{\extracolsep{\fill}}cccccc@{}}
\toprule
Algorithm & P$@$2 & P$@$100 & P$@$500 & P$@$1000 \\
\hline
DeepWalk-emb & \textbf{1} & 0.597 & 0.305 & 0.197  \\
Common Neighbor & \textbf{1} & 0.960 & 0.541 & 0.356 \\
\hline
iUAGA-S  & \textbf{1} & \textbf{1} & \textbf{0.923} & \textbf{0.857} \\
\bottomrule
\end{tabular*}
\end{table}
\end{center}
\begin{figure}[!t]
\centering
\subfigure[Before alignment]{
\label{fig33}
\includegraphics[width=2.8in]{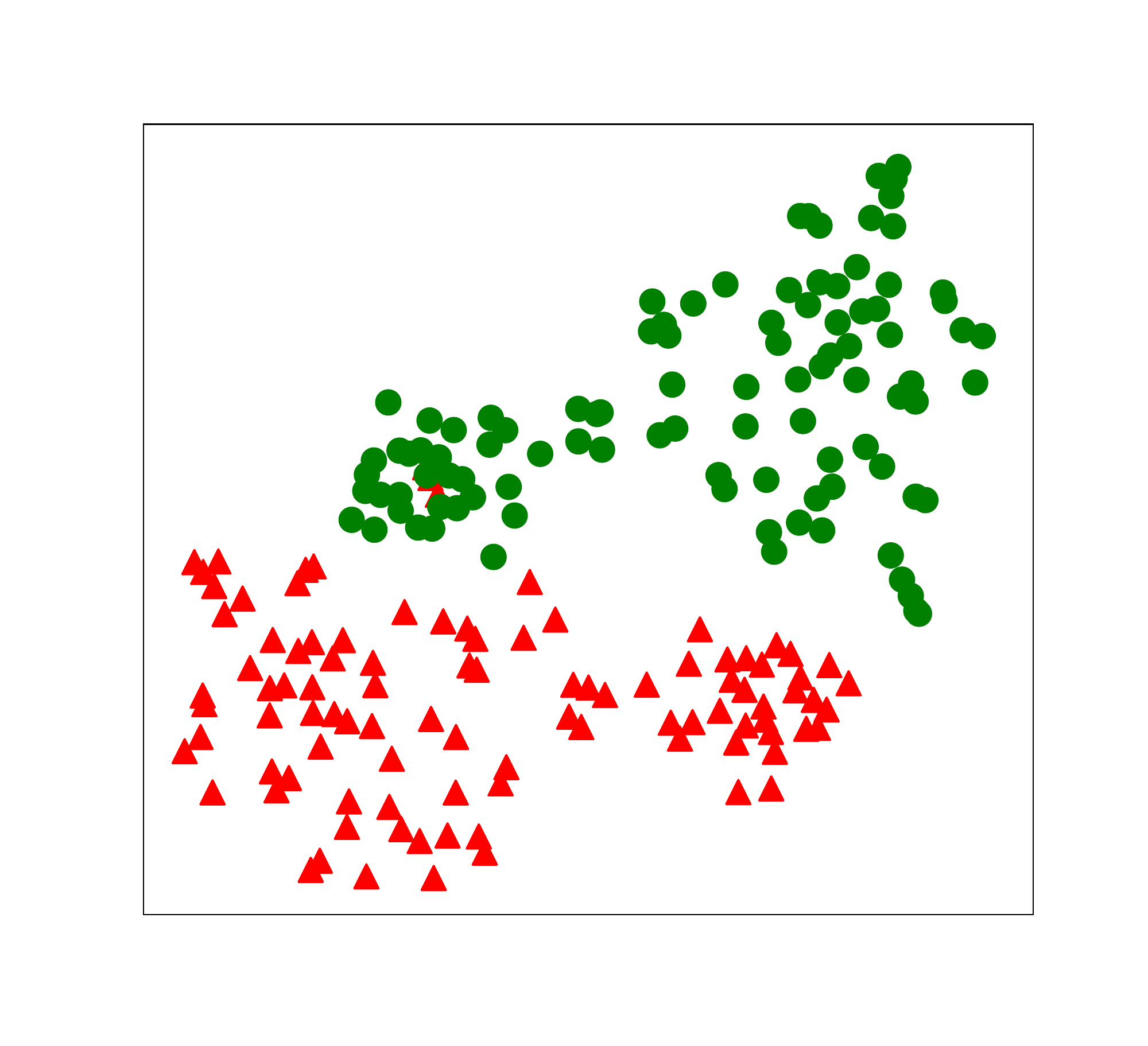}
}
\subfigure[After alignment]{
\label{fig34}
\includegraphics[width=2.8in]{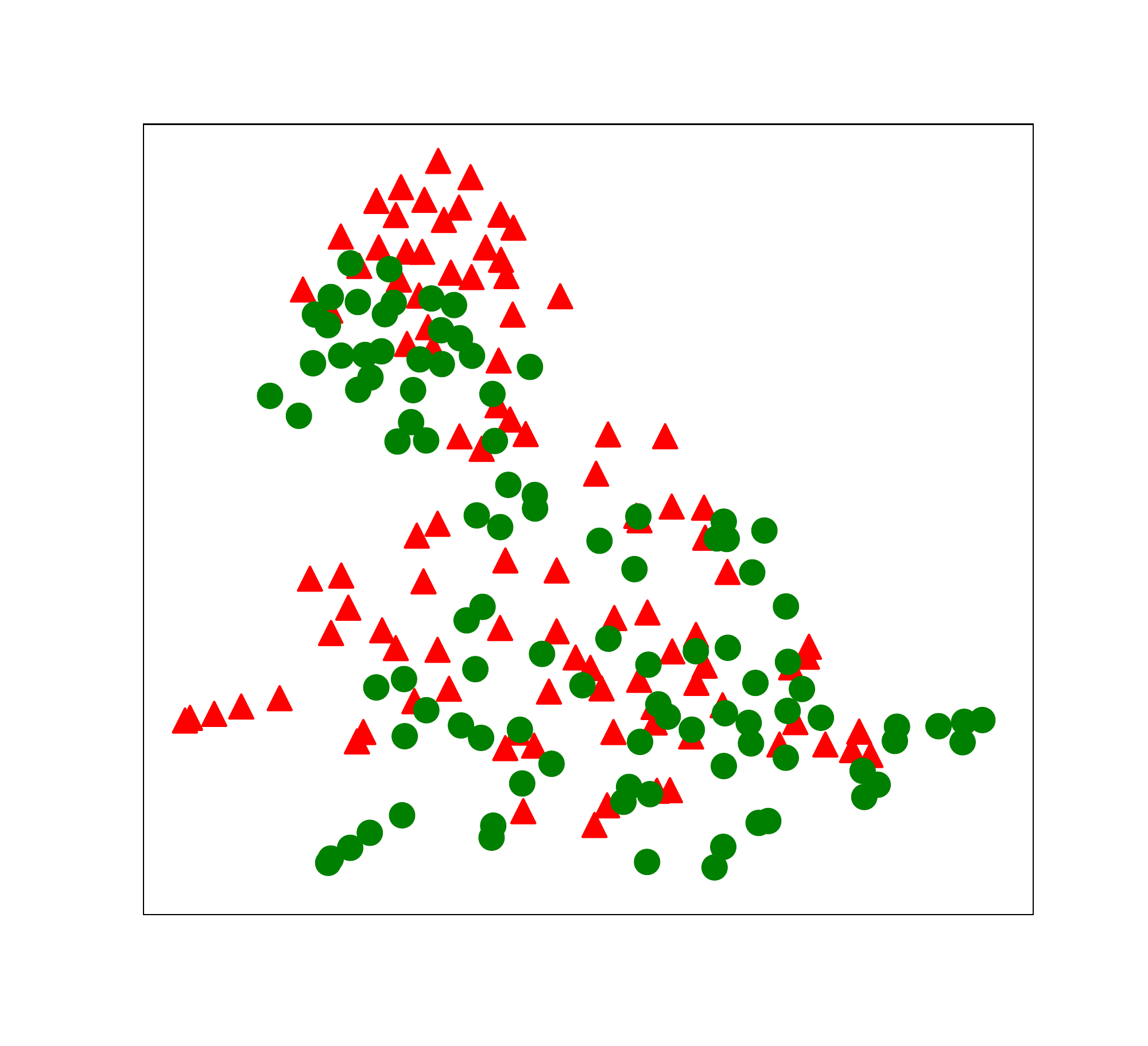}}
\caption{The t-SNE visualization of the source and target graph embedding learned by DeepWalk~(before alignment) and UAGA (after alignment), respectively.}
\label{fig.alignmentres}
\end{figure}

\vspace{-6ex}
\subsection{Case Studies}
In this part, several case studies are implemented to evaluate the effectiveness of the proposed methods on real-world applications.
\subsubsection{Anchor Link Inference between Different Social Graphs}
The results of anchor link inference between different social graphs are reported on Table~\ref{table2}. In terms of the $Precision@N$ (for $N$=10, 20, 30), our proposed approaches significantly outperforms the unsupervised baselines. For example, Adv-Refine-CGSS outperforms Degree by 0.298 in terms of $precision@10$ on $S4$. In particular, we observe that Adv-Refine-CGSS shows comparable or even better performances on most cases compared with Procrustes-CGSS, without requiring any cross-graph annotated data.
\subsubsection{Link Prediction}
We utilize $G_s$ ($\lambda_e=0.9$) of $S1$ as the testing graph for the link prediction and want to evaluate the effectiveness of iUAGA in terms of improving the link prediction accuracy. We follow the same evaluation metric (\emph{i.e.} $precision@k$) with \cite{wang2016structural}. We consider the following comparisons.
\begin{itemize}
    \item \textbf{DeepWalk-emb}, which utilizes DeepWalk to learn the embedding of $G_s$ and then use the obtained representations to predict the unobserved user links;
    \item \textbf{Common Neighbor}~\cite{liben2007link}, which only uses the number of common neighbors to measure the similarity between nodes for user links prediction;
    \item \textbf{iUAGA-S}, which utilizes the iUAGA method to add pseudo user links to $G_s$ by leveraging knowledge from $G_t$ (its counterpart in $S1$) and then use the extended $G_s$ to learn the graph embedding.
\end{itemize}
The results are shown in Table~\ref{table4}, we can see that iUAGA-S significantly outperforms the other baselines, \emph{e.g.} 0.541~$\rightarrow$~\textbf{0.923} ($P@500$) and 0.356~$\rightarrow$~\textbf{0.857} ($P@1000$). This confirms that iUAGA, which incrementally pseudo-labels the unobserved user links with high accuracy, is capable of improving the embedding quality, thus resulting in a more accurate link prediction.

\subsubsection{Embedding Visualization}
In this part, we utilize t-SNE~\cite{donahue2014decaf} to visualize the embedding features of both the source graph (randomly extract 100 nodes together with 1268 edges from $G_s$ of $S1$) and the target graph (their counterparts in $G_t$ of $S1$) before and after graph alignment by UAGA, respectively. The results are presented in Figure~\ref{fig33}-\ref{fig34}, where we can observe that the embedding features become much more domain-invariant after graph alignment. The justification is that the feature discrepancy is significantly reduced in the embedding space of our proposed UAGA.

\section{Conclusions}
\label{sec:conclusions}
In this paper, we study the unsupervised graph alignment problem. We propose a UAGA framework, which learns the source and target embedding spaces in an unsupervised embedding manner, and then attempts to align the two spaces via adversarial training, followed by a refinement procedure. The proposed methods do not require any cross-graph annotated data. Furthermore, we extend our UAGA to iUAGA, which we iteratively reveal the unobserved user links based on the pseudo anchor links. Thus it can be used to further improve the embedding quality and the final alignment accuracy. Our proposed approaches have achieved better performance than several unsupervised methods, and received comparable or even better performances than supervised methods on real-world data, which clearly demonstrates the effectiveness of our newly proposed approaches for unsupervised graph alignment.

\balance
\bibliographystyle{ieeetr}
\bibliography{main}

\begin{thebibliography}{10}

\bibitem{dong2012link}
Y.~Dong, J.~Tang, S.~Wu, J.~Tian, N.~V. Chawla, J.~Rao, and H.~Cao, ``Link
  prediction and recommendation across heterogeneous social networks,'' in {\em
  2012 IEEE 12th International Conference on Data Mining}, pp.~181--190, IEEE,
  2012.

\bibitem{zhang2014meta}
J.~Zhang, P.~S. Yu, and Z.-H. Zhou, ``Meta-path based multi-network collective
  link prediction,'' in {\em Proceedings of the 20th ACM SIGKDD international
  conference on Knowledge discovery and data mining}, pp.~1286--1295, ACM,
  2014.

\bibitem{hu2013personalized}
L.~Hu, J.~Cao, G.~Xu, L.~Cao, Z.~Gu, and C.~Zhu, ``Personalized recommendation
  via cross-domain triadic factorization,'' in {\em Proceedings of the 22nd
  international conference on World Wide Web}, pp.~595--606, ACM, 2013.

\bibitem{li2014matching}
C.-Y. Li and S.-D. Lin, ``Matching users and items across domains to improve
  the recommendation quality,'' in {\em Proceedings of the 20th ACM SIGKDD
  international conference on Knowledge discovery and data mining},
  pp.~801--810, ACM, 2014.

\bibitem{yacsar2018iterative}
A.~Ya{\c{s}}ar and {\"U}.~V. {\c{C}}ataly{\"u}rek, ``An iterative global
  structure-assisted labeled network aligner,'' in {\em Proceedings of the 24nd
  ACM SIGKDD International Conference on Knowledge Discovery and Data Mining},
  pp.~2614--2623, ACM, 2018.

\bibitem{huang2018adaptive}
W.~Huang, T.~Zhang, Y.~Rong, and J.~Huang, ``Adaptive sampling towards fast
  graph representation learning,'' in {\em Advances in Neural Information
  Processing Systems}, pp.~4558--4567, 2018.

\bibitem{li2019semi}
J.~Li, Y.~Rong, H.~Cheng, H.~Meng, W.~Huang, and J.~Huang, ``Semi-supervised
  graph classification: A hierarchical graph perspective,'' 2019.

\bibitem{singh2008global}
R.~Singh, J.~Xu, and B.~Berger, ``Global alignment of multiple protein
  interaction networks with application to functional orthology detection,''
  {\em Proceedings of the National Academy of Sciences}, 2008.

\bibitem{bayati2009algorithms}
M.~Bayati, M.~Gerritsen, D.~F. Gleich, A.~Saberi, and Y.~Wang, ``Algorithms for
  large, sparse network alignment problems,'' in {\em Data Mining, 2009.
  ICDM'09. Ninth IEEE International Conference on}, pp.~705--710, IEEE, 2009.

\bibitem{koutra2013big}
D.~Koutra, H.~Tong, and D.~Lubensky, ``Big-align: Fast bipartite graph
  alignment,'' in {\em 2013 IEEE 13th International Conference on Data Mining},
  pp.~389--398, IEEE, 2013.

\bibitem{zhang2016final}
S.~Zhang and H.~Tong, ``Final: Fast attributed network alignment,'' in {\em
  Proceedings of the 22nd ACM SIGKDD International Conference on Knowledge
  Discovery and Data Mining}, pp.~1345--1354, ACM, 2016.

\bibitem{kong2013inferring}
X.~Kong, J.~Zhang, and P.~S. Yu, ``Inferring anchor links across multiple
  heterogeneous social networks,'' in {\em Proceedings of the 22nd ACM
  international conference on Information \& Knowledge Management},
  pp.~179--188, ACM, 2013.

\bibitem{zafarani2013connecting}
R.~Zafarani and H.~Liu, ``Connecting users across social media sites: a
  behavioral-modeling approach,'' in {\em Proceedings of the 19th ACM SIGKDD
  international conference on Knowledge discovery and data mining}, pp.~41--49,
  ACM, 2013.

\bibitem{tan2014mapping}
S.~Tan, Z.~Guan, D.~Cai, X.~Qin, J.~Bu, and C.~Chen, ``Mapping users across
  networks by manifold alignment on hypergraph.,'' in {\em AAAI}, vol.~14,
  pp.~159--165, 2014.

\bibitem{liu2016aligning}
L.~Liu, W.~K. Cheung, X.~Li, and L.~Liao, ``Aligning users across social
  networks using network embedding.,'' in {\em IJCAI}, pp.~1774--1780, 2016.

\bibitem{man2016predict}
T.~Man, H.~Shen, S.~Liu, X.~Jin, and X.~Cheng, ``Predict anchor links across
  social networks via an embedding approach.,'' in {\em IJCAI}, vol.~16,
  pp.~1823--1829, 2016.

\bibitem{schonemann1966generalized}
P.~H. Sch{\"o}nemann, ``A generalized solution of the orthogonal procrustes
  problem,'' {\em Psychometrika}, vol.~31, no.~1, pp.~1--10, 1966.

\bibitem{klau2009new}
G.~W. Klau, ``A new graph-based method for pairwise global network alignment,''
  {\em BMC bioinformatics}, vol.~10, no.~1, p.~S59, 2009.

\bibitem{liu2014hydra}
S.~Liu, S.~Wang, F.~Zhu, J.~Zhang, and R.~Krishnan, ``Hydra: Large-scale social
  identity linkage via heterogeneous behavior modeling,'' in {\em Proceedings
  of the 2014 ACM SIGMOD international conference on Management of data},
  pp.~51--62, ACM, 2014.

\bibitem{zhong2012comsoc}
E.~Zhong, W.~Fan, J.~Wang, L.~Xiao, and Y.~Li, ``Comsoc: adaptive transfer of
  user behaviors over composite social network,'' in {\em Proceedings of the
  18th ACM SIGKDD international conference on Knowledge discovery and data
  mining}, pp.~696--704, ACM, 2012.

\bibitem{zhang2015multiple}
J.~Zhang and S.~Y. Philip, ``Multiple anonymized social networks alignment,''
  in {\em Data Mining (ICDM), 2015 IEEE International Conference on},
  pp.~599--608, IEEE, 2015.

\bibitem{wang2016structural}
D.~Wang, P.~Cui, and W.~Zhu, ``Structural deep network embedding,'' in {\em
  Proceedings of the 22nd ACM SIGKDD international conference on Knowledge
  discovery and data mining}, pp.~1225--1234, ACM, 2016.

\bibitem{qiu2018network}
J.~Qiu, Y.~Dong, H.~Ma, J.~Li, K.~Wang, and J.~Tang, ``Network embedding as
  matrix factorization: Unifying deepwalk, line, pte, and node2vec,'' in {\em
  Proceedings of the Eleventh ACM International Conference on Web Search and
  Data Mining}, pp.~459--467, ACM, 2018.

\bibitem{kazemi2018simple}
S.~M. Kazemi and D.~Poole, ``Simple embedding for link prediction in knowledge
  graphs,'' {\em arXiv preprint arXiv:1802.04868}, 2018.

\bibitem{perozzi2014deepwalk}
B.~Perozzi, R.~Al-Rfou, and S.~Skiena, ``Deepwalk: Online learning of social
  representations,'' in {\em Proceedings of the 20th ACM SIGKDD international
  conference on Knowledge discovery and data mining}, pp.~701--710, ACM, 2014.

\bibitem{tang2015line}
J.~Tang, M.~Qu, M.~Wang, M.~Zhang, J.~Yan, and Q.~Mei, ``Line: Large-scale
  information network embedding,'' in {\em Proceedings of the 24th
  International Conference on World Wide Web}, pp.~1067--1077, International
  World Wide Web Conferences Steering Committee, 2015.

\bibitem{grover2016node2vec}
A.~Grover and J.~Leskovec, ``node2vec: Scalable feature learning for
  networks,'' in {\em Proceedings of the 22nd ACM SIGKDD international
  conference on Knowledge discovery and data mining}, pp.~855--864, ACM, 2016.

\bibitem{mikolov2013distributed}
T.~Mikolov, I.~Sutskever, K.~Chen, G.~S. Corrado, and J.~Dean, ``Distributed
  representations of words and phrases and their compositionality,'' in {\em
  Advances in neural information processing systems}, pp.~3111--3119, 2013.

\bibitem{mikolov2013efficient}
T.~Mikolov, K.~Chen, G.~Corrado, and J.~Dean, ``Efficient estimation of word
  representations in vector space,'' {\em arXiv preprint arXiv:1301.3781},
  2013.

\bibitem{goodfellow2014generative}
I.~Goodfellow, J.~Pouget-Abadie, M.~Mirza, B.~Xu, D.~Warde-Farley, S.~Ozair,
  A.~Courville, and Y.~Bengio, ``Generative adversarial nets,'' in {\em
  Advances in neural information processing systems}, pp.~2672--2680, 2014.

\bibitem{dinu2014improving}
G.~Dinu, A.~Lazaridou, and M.~Baroni, ``Improving zero-shot learning by
  mitigating the hubness problem,'' {\em arXiv preprint arXiv:1412.6568}, 2014.

\bibitem{johnson2017billion}
J.~Johnson, M.~Douze, and H.~J{\'e}gou, ``Billion-scale similarity search with
  gpus,'' {\em arXiv preprint arXiv:1702.08734}, 2017.

\bibitem{smith2017offline}
S.~L. Smith, D.~H. Turban, S.~Hamblin, and N.~Y. Hammerla, ``Offline bilingual
  word vectors, orthogonal transformations and the inverted softmax,'' {\em
  arXiv preprint arXiv:1702.03859}, 2017.

\bibitem{cisse2017parseval}
M.~Cisse, P.~Bojanowski, E.~Grave, Y.~Dauphin, and N.~Usunier, ``Parseval
  networks: Improving robustness to adversarial examples,'' in {\em Proceedings
  of the 34th International Conference on Machine Learning-Volume 70},
  pp.~854--863, JMLR. org, 2017.

\bibitem{liben2007link}
D.~Liben-Nowell and J.~Kleinberg, ``The link-prediction problem for social
  networks,'' {\em Journal of the American society for information science and
  technology}, vol.~58, no.~7, pp.~1019--1031, 2007.

\bibitem{donahue2014decaf}
J.~Donahue, Y.~Jia, O.~Vinyals, J.~Hoffman, N.~Zhang, E.~Tzeng, and T.~Darrell,
  ``Decaf: A deep convolutional activation feature for generic visual
  recognition,'' in {\em International conference on machine learning},
  pp.~647--655, 2014.

\end{thebibliography}
\end{document}